\documentclass[aps, reprint, showpacs, longbibliography, superscriptaddress]{revtex4-1}
\usepackage{color, graphicx}
\usepackage{amssymb,amsmath, amsfonts, bm}
\usepackage[colorlinks,citecolor=blue, urlcolor=blue, linkcolor=blue]{hyperref}
\usepackage{bbold}

\begin{document}
\title{Efficient implementation of the parquet equations -- role of the reducible vertex function and its kernel approximation}  

\author{Gang Li}
\email[Correspondence and requests for materials should be addressed to: ]{gangli.phy@gmail.com}
\affiliation{\mbox{Institute of Solid State Physics, Vienna University of Technology, A-1040 Vienna, Austria}}  
\author{Nils Wentzell}
\affiliation{\mbox{Institute of Solid State Physics, Vienna University of Technology, A-1040 Vienna, Austria}}  
\affiliation{Institut f\"ur Theoretische Physik and CQ Center for Collective Quantum Phenomena,
  Universit\"at T\"ubingen, Auf der Morgenstelle 14, 72076 T\"ubingen, Germany}
\author{Petra Pudleiner}
\affiliation{\mbox{Institute of Solid State Physics, Vienna University of Technology, A-1040 Vienna, Austria}}  
\author{Patrik Thunstr\"om}
\affiliation{\mbox{Institute of Solid State Physics, Vienna University of Technology, A-1040 Vienna, Austria}}  
\author{Karsten Held}
\affiliation{\mbox{Institute of Solid State Physics, Vienna University of Technology, A-1040 Vienna, Austria}}  


\begin{abstract}
We present an efficient implementation of the parquet formalism which respects the asymptotic structure of the vertex functions at both single- and two-particle levels in momentum- and frequency-space. 
We identify the two-particle reducible vertex as the core function which is essential for the construction of the other vertex functions. 
This observation stimulates us to consider a two-level parameter-reduction for this function to simplify the solution of the parquet equations. 
The resulting functions, which depend on fewer arguments, are coined ``kernel functions". With the use of the ``kernel functions", the open boundary of various vertex functions in the Matsubara-frequency space can be faithfully satisfied. We justify our implementation by accurately reproducing 
the dynamical mean-field theory results from momentum-independent parquet calculations. 
The high-frequency asymptotics of the single-particle self-energy and the two-particle vertex are correctly reproduced, which turns out to be essential for the self-consistent determination of the parquet solutions. 
The current implementation is also feasible for the dynamical vertex approximation.
\end{abstract}

\maketitle
\section{Introduction}
Strong electronic correlations lead to the arguably most fascinating and least understood phenomena in solid state physics
such as the breakdown of Landau's \cite{landau1957theory, Lifshitz1980} Fermi liquid theory and high temperature superconductivity~\cite{High-Tc}.
However, solving the correlated electron problem poses a great challenge to theoretical physics, since the competition between interaction and kinetic energy prohibits a simple perturbative treatment of such many-body systems. The minimal model covering this competition between localizing and delocalizing electrons is the Hubbard model~\cite{Hubbard}. 
Only in the special cases where one energy-scale dominates, weak-~\cite{PhysRev.82.625, doi:10.1142/S021797929100016X, PhysRev.85.338,PhysRev.92.609,PhysRevB.50.14016,PhysRevB.55.2122} or strong-coupling~\cite{strongcoupling_1,PhysRevB.53.2691,PhysRevLett.103.086403} perturbative treatments are actually reliable. 
 
Many of these perturbative approximations are functional-derivable, which is a key criterion that Baym and Kadanoff~\cite{PhysRev.124.287} discovered for a many-body theory to be conservative. 
They found that for any functional that is derivable with respect to the single-particle propagator, the resulting self-energy function and the Green's function satisfy the continuity equations. 
The central object in these conservative theories are the single-particle self-energy, which, in the Baym-Kadanoff formalism, can be calculated self-consistently. 
An alternative to the Baym-Kadanoff formalism, that is self-consistent also at the two-particle level, was developed by Landau, Dominicis and Martin~\cite{abrikosov1954elimination, Dominicis_Martin_1, Dominicis_Martin_2}, which is referred to as the parquet formalism. 
The central object in this theory is the two-particle vertex functions, from which the single-particle self-energy can be self-consistently calculated. 
The parquet formalism has the self-consistency at both single- and two-particle levels built in, which by construction can be better than the the Baym-Kadanoff theorem in this respect.
However, unlike the Baym-Kadanoff theorem, the parquet equations do not explicitly guarantee to satisfy the conservation laws, such as the continuity equations. 

The generalization of the self-consistency from the single-particle to the two-particle level is essential to describe the behavior of individual particles and their collective excitations on an equal footing. 
One example of such complexity is the spin-fluctuation-mediated pairing interaction in the cuprate superconductors~\cite{PhysRevLett.62.961, Scalapino_1}. 
To answer how two individual particles form a Cooper pair in the particle-particle channel requires the knowledge of the spin fluctuations in the particle-hole channel.  
In this problem, both the single-particle delocalization and the two-particle excitations need to be determined simutaneously, which calls for a theory with self-consistency at both the single- and the two-particle level. 
But this is not limited to this particular example. 
In general, for any collective order that arises from the competition between different fluctuations and low-energy excitations, one needs a theory like the parquet formalism that satisfies the self-consistency at both single- and two-particle levels. 
However, the application of the parquet equations so far has been  limited to only a few cases~\cite{PhysRevB.46.8050, PhysRevB.47.8851, PhysRevB.75.165108, PhysRevB.83.035114,Chen1992311,PhysRevB.43.8044,PhysRevE.80.046706,PhysRevE.87.013311,PhysRevB.91.115115}. 
The main obstacle for the parquet equations from being widely applied is the numerical feasibility.  
The two-particle vertex depends on three independent arguments, each of which  consists of both momentum and frequency. 
Even in the $SU(2)$ symmetric case, solving the four coupled parquet equations for a reasonably large system at low-temperature is still numerically very challenging. 
Here, the difficulty does not only concern the storage of the large two-particle vertices. Of even more concern is how to actually preserve the asymptotic structure of the single-particle self-energy and the two-particle vertices simultaneously during the calculation. 
Due to the fact that the parquet self-consistency is performed on both the single- and the two-particle level, the truncation of the two-particle vertex structures will unavoidably result in a wrong evaluation of the single-particle self-energy, and vice versa.  
In a consistent solution of the parquet equations, the correct self-energy as well as all vertex functions should be simultaneously obtained at convergence. 

In this paper, we present a new and efficient implementation of the parquet equations which satisfies a number of important conditions. The prime interest of our implementation is to correctly reproduce the asymptotics for the single-particle self-energy and the two-particle vertex functions at each self-consistent step by employing a precise inner and an asymptotic outer frequency window, which ensures that the converged solutions are consistent and asymptoticlly correct.   

The paper is organized as follows: For completeness, we introduce the necessary notations for the single- and two-particle vertex in Sec.~\ref{Sec:PA_formalism}. We also briefly derive the corresponding formalism for the parquet equations and the self-energy in this notation. For the readers who are familiar with the parquet formalism and are only interested in the detailed implementation, this part can be safely skipped. 
In Sec.~\ref{Sec:PA_solution}, which is the main part of this paper, we present our philosophy for solving the parquet equations. 
In accordance with previous findings~\cite{PhysRevB.86.125114} we identify the dominant structures in the two-particle vertex. We reduce their complexity by focussing only on the parts reducible in a specific channel, motivating our two-level kernel approximation.
In Sec.~\ref{Sec:PA_results}, we solve the Anderson impurity model and a $2\times2$ cluster within the full parquet and the dynamical vertex approximation, respectively. For the former we have the exact results from the dynamical mean-field theory (DMFT)~\cite{RevModPhys.68.13} which in turn justifies our implementation of the parquet equations. An excellent agreement is achieved at both the single-and two-particle levels. 
A summary and outlook is provided in Sec.~\ref{Sec:PA_summary}.

\section{Solution of the parquet equations}\label{Sec:PA_solution}
The parquet equation is a classification of the full vertex $F$ 
into the (two-particle) fully irreducible contributions $\Lambda$ and the reducible contributions in
 the particle-hole ($\Phi$), the transversal particle-hole (follows by symmetry) and the particle-particle channel ($\Psi$). Employing the $SU(2)$ symmetry, one can decouple their spin components into density ($d$)/magnetic ($m$) and singlet ($s$)/triplet ($t$) channel, respectively. 
In these four channels, the parquet equation reads:
\begin{subequations}\label{PA_F}
\begin{align}
F_{d/m}^{k,k^{\prime}}(q)=\Lambda_{d/m}^{k,k^{\prime}}(q) &+ \Phi^{k,k^{\prime}}_{d/m}(q) +c_{1}^{d/m}\cdot\Phi_{d}^{k,k+q}(k^{\prime}-k)\nonumber\\
&+c_{2}^{d/m}\cdot\Phi_{m}^{k,k+q}(k^{\prime}-k)\nonumber\\
&+c_{3}^{d/m}\cdot\Psi_{s}^{k,k^{\prime}}(k+k^{\prime}+q)\nonumber\\
&+c_{4}^{d/m}\cdot\Psi_{t}^{k,k^{\prime}}(k+k^{\prime}+q)\;;\label{PA_F_dm}\\
F_{s/t}^{k,k^{\prime}}(q)=\Lambda_{s/t}^{k,k^{\prime}}(q) &+ \Psi^{k,k^{\prime}}_{s/t}(q) 
+c_{1}^{s/t}\cdot\Phi_{d}^{k,q-k^{\prime}}(k^{\prime}-k)\nonumber\\
 &+c_{2}^{s/t}\cdot\Phi_{m}^{k,q-k^{\prime}}(k^{\prime}-k)\nonumber\\
& +c_{3}^{s/t}\cdot\Phi_{d}^{k,k^{\prime}}(q-k-k^{\prime})\nonumber\\
& +c_{4}^{s/t}\cdot\Phi_{m}^{k,k^{\prime}}(q-k-k^{\prime})\:.\label{PA_F_st}  
\end{align}
\end{subequations} 
$k=(\mathbf{k},i \nu)$ is a compound index consisting of  wave vector $\mathbf{k}$ and Matsubara frequency $i \nu$. 
The coefficients $c_{1\cdots4}^{d/m/s/t}$ take different values in the four different channels.
We only briefly list here the necessary equations for the convenience of the discussions in the main part of the paper, more detailed notations and derivations can be found in Appendix~\ref{Sec:PA_formalism}.
In Eq.~(\ref{PA_F}), the reducible contributions are given by the  Bethe-Salpeter equation (BSE) in the four channels formally as  $\Phi/\Psi=\Gamma  G G F$ [Eq. (\ref{PA_F_Phi})]. 
Here, $\Gamma$ is the irreducible vertex in the given channel which contains 
the reducible contributions from the other channels and the fully irreducible $\Lambda$, see  Eqs. (\ref{PA_Gamma_dm}) and (\ref{PA_Gamma_st}). 
The self-consistency at the single- and two-particle level are synchronized by means of the self-energy which depends on the resulting two-particle vertex as shown in the Schwinger-Dyson equation of motion (\ref{PA_Sigma}). 

Given the fully irreducible vertex $\Lambda$, 
 the parquet formalism provides a set of five exact equations [(\ref{PA_F}),  (\ref{PA_F_Phi}),  (\ref{PA_Gamma_dm})/(\ref{PA_Gamma_st}), (\ref{PA_Sigma}), (\ref{PA_Dyson})] which  can be solved for the  five unknowns [$F$, $\Phi/\Psi$, $\Gamma$, $G$, $\Sigma$] (where the former three equations and vertices consist of four channels each).
Hence, if we know the exact $\Lambda$, we can calculate all physical, one- and two-particle, quantities exactly. 
However, since the exact  $\Lambda$ of the Hubbard model is not known, we need to make approximations.
In the parquet approximation (PA)~\cite{doi:10.1142/S021797929100016X,Bickers-Review}, $\Lambda\sim U$ is taken; A more sophisticated approximation that takes into account all local fully irreducible diagrams is referred to as the dynamical vertex approximation~\cite{PhysRevB.75.045118, Held-review}.

In this paper, we mainly discuss two problems that are practically unavoidable in solving the parquet equations, which are of critical importance for keeping the self-consistency in the single- and two-particle levels simultaneously. 

The first problem arises due to the finite numbers of Matsubara frequencies that are available in the calculations. 
Each vertex in the parquet equation depends on three independent arguments $k$, $k^{\prime}$ and $q$,  which take arbitrary values in $(-\infty, \infty)$. 
In practice, a finite cutoff $a$ has to be introduced. A consequence of this cut-off is that after each self-consistency step the interval on which the vertex is known shrinks. 
This can be seen as follows: 
Take equation (\ref{PA_F_dm}) as an example and suppose $k$, $k^{\prime}$ and $q$ to take values in $[-a, a]$. For calculating the the right-hand-side of (\ref{PA_F_dm}), we would need the solutions of $\Phi_{d/m}^{k,k+q}(k^{\prime}-k)$ in $[-2a, 2a]$, and $\Psi_{s/t}^{k,k^{\prime}}(k+k^{\prime}-q)$ in $[-3a, 3a]$. 
Assuming that $\Phi_{d/m}^{k,k^{\prime}}(q)$ and $\Psi_{s/t}^{k,k^{\prime}}(q)$  are only available in $[-a, a]$, $F_{d/m}^{k,k^{\prime}}(q)$ can then be calculated only in the smaller interval $[-a/3, a/3]$. 
Such a boundary issue only exists in the Matsubara frequency space. In momentum space, the periodic boundary condition can be applied whenever $\mathbf{k}^{\prime}-\mathbf{k}$ or $\mathbf{k}+\mathbf{k}^{\prime}+\mathbf{q}$ exceed the finite parameter range. 
However, none of the vertex functions is periodically dependent on the Matsubara frequencies $i\nu, i\nu^{\prime}$ and $i\omega$~\cite{, PhysRevE.87.013311}.  
As a result, there exists two different parameter spaces for the vertex functions, {\it i.e.} in the bigger space ($[-a, a]$) $\Phi_{d/m}^{k,k^{\prime}}(q)$ and $\Psi_{s/t}^{k,k^{\prime}}(q)$ are known, while through the parquet equations
$F_{d/m/s/t}^{k,k^{\prime}}(q)$ can be determined only in a smaller parameter space ($[-a/3, a/3]$).  

\begin{figure}[tbp]
\centering
\includegraphics[width=\linewidth]{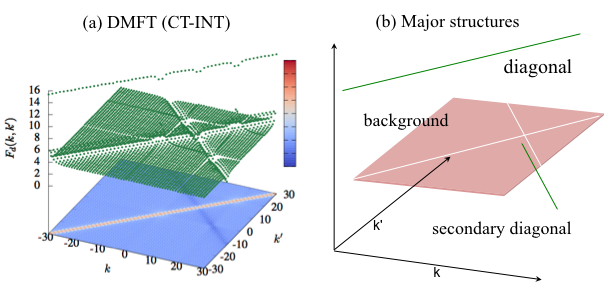}
\caption{(a) The green dots are the full vertex $F_{d}^{k,k^{\prime}}(q)$  for a fixed value of $q$ calculated from the DMFT (CT-INT) at $\beta t=2$ and $U/t=4$ on a square lattice. The bottom shows the intensity of $F_{d}^{k,k^{\prime}}(q)$, which illustrate three major structures of the vertex functions. 
That is background, diagonal and secondary diagonal components as illustrated in (b).}
\label{Fig:vertex_structure}
\end{figure}

The second problem is related to the finite frequency parameter range as well. 
To evaluate the self-energy function in Eq.~(\ref{PA_Sigma}), a sum over the two internal arguments $k^{\prime}$ and $q$ has to be carried out. 
An example of the vertex functions $F_{d}^{\nu,\nu^{\prime}}(\omega)$ is shown in Fig.~\ref{Fig:vertex_structure}. As was already observed in Ref.~\cite{PhysRevB.86.125114, Georgphdthesis}, $F_{d}^{\nu,\nu^{\prime}}(\omega)$  has structures that span the whole Matsubara frequency space.
In particular,  they do not decay at the boundary of any given parameter box.  
Thus, a sum over a finite parameter range corresponds to a truncation of these vertex functions at the boundary which can lead to a wrong evaluation of the self-energy function. 

In this paper, we propose a feasible scheme to solve these two problems, improving upon the Matsubara-frequency periodization employed hitherto~\cite{PhysRevE.87.013311}. Our idea is based on the observation of the central role the reducible vertex functions play in the parquet equations, which will be explained in the following. 

\subsection{Two-level kernel approximations}
To satisfy the crossing symmetry explicitly in every self-consistency step, we evaluate the full vertex $F_{d/m/s/t}$ directly from the parquet equations ~\cite{PhysRevE.87.013311}. 
Figure~\ref{Fig:vertex_structure} displays $F_{d}^{k,k^{\prime}}(q)$ as a function of $k$ and $k^{\prime}$ for a fixed $q$.  
The left plot is obtained from a DMFT calculation with the interaction-expansion continuous-time quantum Monte Carlo (CT-INT)~\cite{PhysRevB.72.035122, RevModPhys.83.349} as an impurity solver, thus it represents a numerically exact (up to the statistical errors of the CT-INT) evaluation of the full two-particle vertex for the DMFT impurity.    
We will calculate this vertex in the parquet theory as well, see Sec.~\ref{Sec:PA_results}.  
A detailed analysis of the two-particle vertex function can be found in Ref.~\cite{PhysRevB.86.125114, Georgphdthesis}. 
In the following, we will use the exact results from DMFT as a reference to further show that, among the various two-particle vertex functions,  the reducible vertex is the most important one, which plays the central role in our implementation of the parquet equations. 

The right plot shows a schematic representation of the major structures of the left one.
The full vertex $F_{d/m/s/t}$ can be decomposed into three main parts, {\it i.e.} the background, the diagonal and the secondary diagonal component.  
Fig.~\ref{Fig:vertex_structure} clearly shows that the boundary of the vertex function is not periodic in frequency space, instead all three components extend to infinite values of $k$ and $k^{\prime}$.  
Due to the restricted parameter space available in practical calculations, one has to be careful with the boundary effect on these vertex functions. 
 
 The background is contributed by $\Lambda_{d/m/s/t}^{k,k^{\prime}}(q)$, which is the input for the parquet equation, and is further supplemented by the reducible vertex functions $\Phi_{d/m}^{k,k^{\prime}}(q)$, $\Psi_{s/t}^{k,k^{\prime}}(q)$. 
The diagonal and the secondary diagonal components are predominant for $F^{k,k^{\prime}}_{d/m/s/t}(q)$ with $k=k^{\prime}$ and $k=-k^{\prime}-q$ in the $d/m$-channel, and for $k=k^{\prime}-q$ in the $s/t$-channel, respectively. 
The diagonal and secondary diagonal components are generated, in the parquet equations, by the reducible vertex $\Phi_{d/m}^{k,k+q}(k^{\prime}-k)$, $\Psi_{s/t}^{k,k^{\prime}}(k+k^{\prime}+q)$ in the $d/m$-channel and $\Phi_{d/m}^{k,q-k^{\prime}}(k^{\prime}-k)$, $\Phi_{d/m}^{k,k^{\prime}}(q-k-k^{\prime})$ in the $s/t$-channel, see Eq.~(\ref{PA_F}).
Furthermore, we also notice that these two components only depend significantly on the center of mass momentum and frequency (which is the momentum/frequency in the brackets); the dependence on the other two arguments (the superscript momentum/frequency)  is much weaker, as will be shown in the following. Hence, the reducible vertex can be effectively approximated by a single-argument dependent functions $\tilde{\Phi}_{d/m}(q)$ and $\tilde{\Psi}_{s/t}(q)$, which we call kernel functions. 
The approximation of replacing the three-argument dependent reducible vertex with a single-$q$ dependent kernel function, {\it i.e.},  $\Phi_{d/m}^{k,k+q}(k^{\prime}-k)  \approx \tilde{\Phi}_{d/m}(\tilde{q}=k^{\prime}-k)$ {\it etc.}, is called the first-level kernel approximation. 
We coin it ``kernel approximation" since, on the one hand, $\tilde{\Phi}(q)$ contains the most essential, {\it i.e.}, core or ``kernel", information of $\Phi^{k,k^{\prime}}(q)$. 
On the  other hand we use this term since, mathematically, the kernel of our mapping ${\cal F}: q,k,k' \rightarrow q$ defines classes of equivalent frequency triples, whose reducible vertex $\Phi^{k,k^{\prime}}(q)$ is (approximatively) the same, {\it i.e.},  $\tilde\Phi(q)$.
The parameter-reduction of the reducible vertex functions, {\it i.e.} the kernel approximation,  will greatly simplify our implementation of the parquet equations. 
Let us emphasize that we only employ the kernel approximation when the Matsubara frequency is outside the interval $[-a,a]$ in which the vertex is known explicitly.
We also note that a parametrization related to the first-level kernel approximation is used in a different context: Karrasch {\it et al.} use a sum of single frequency full vertex functions for the functional renormalization group calculations~\cite{0953-8984-20-34-345205}, where however this parametrization is employed for all frequencies.

\begin{figure}[tbp]
\centering
\includegraphics[width=\linewidth]{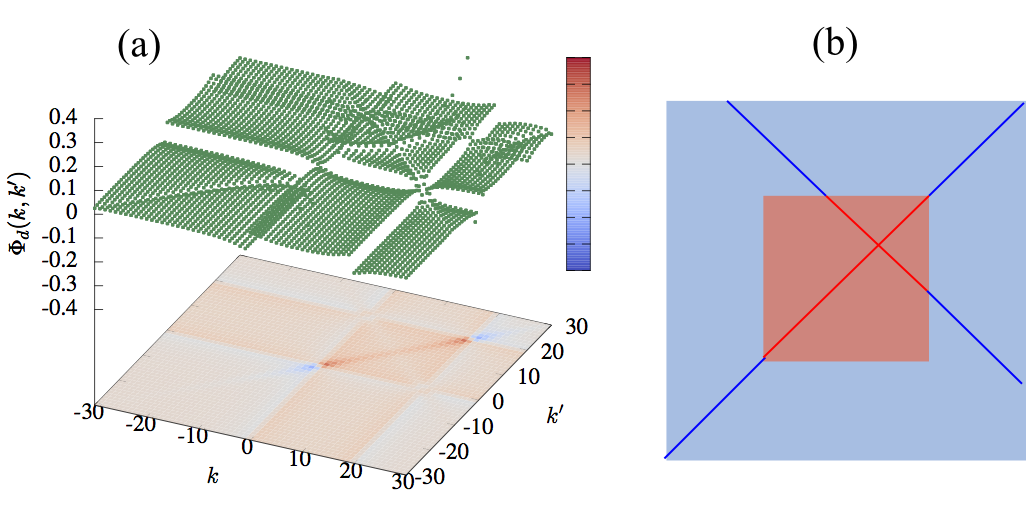}
\caption{(a) Reducible vertex in the density channel calculated from the DMFT (CT-INT) for the same parameter as in Fig.~\ref{Fig:vertex_structure}.
(b) Schematic illustration of our philosophy of the kernel-approximation(s) for solving the open boundary issue in the parquet equations, see main text for more details.}
\label{Fig:reducible_vertex}
\end{figure}

We verify the simple structure of the reducible vertex functions from a DMFT calculation in Fig.~\ref{Fig:reducible_vertex}(a), where $\Phi_{d}^{k,k^{\prime}}(q)$ is displayed as a function of $k$ and $k^{\prime}$ for a fixed transfer frequency $q=i\omega=-i40\pi/\beta$.  
First of all, we notice that the overall amplitude of the reducible vertex function for the given parameters is much smaller than that of the full vertex shown in  Fig.~\ref{Fig:vertex_structure} for the same parameters. 
Compared to Fig.~\ref{Fig:vertex_structure}, the reducible vertex can rather be viewed as a flat plane.  
Secondly, the detailed structure of the reducible vertex is found to consist of only two main parts, {\it i.e.}, a constant background and two crossing stripes.  
The first-level kernel approximation discussed above corresponds to considering only the constant background.  
In practice, as the first-level kernel function $\tilde{\Phi}_{d/m}(q)$ [$\tilde{\Psi}_{s/t}(q)$] we take for every $q$ the value of $\Phi_{d/m}^{k,k^{\prime}}(q)$  [$\Psi_{s/t}^{k,k^{\prime}}(q)$] at this $q$ and a $k$, $k ^{\prime}$ that is far away from the diagonal components and the stripes in  Fig.~\ref{Fig:reducible_vertex}(b).
There is certain freedom in this choice, that is yet to be further investigated. 

For an intuitive understanding of this approximation, let us examine the first iteration of the PA. 
Here, $\Lambda_{d/m/s/t}^{k,k^{\prime}}(q)$, $F_{d/m/s/t}^{k,k^{\prime}}(q)$ and $\Gamma_{d/m/s/t}^{k,k^{\prime}}(q)$ are simply taken as $(U, -U, 2U, 0)$. 
From Eq.~(\ref{PA_F_Phi}), we learn
\begin{eqnarray}
\Phi_{d/m}^{k,k^{\prime}}(q) &=& \frac{U^{2}}{\beta N}\sum_{k^{\prime\prime}}G(k^{\prime\prime})G(k^{\prime\prime}+q)\;,\nonumber\\
\Phi_{s}^{k,k^{\prime}}(q) &=& -\frac{2U^{2}}{\beta N}\sum_{k^{\prime\prime}}G(k^{\prime\prime})G(q-k^{\prime\prime})\;,\nonumber\\
\Phi_{t}^{k,k^{\prime}}(q)&=& 0\;.
\end{eqnarray}
which depend on $q$ only. For any given $q$, $\Phi_{d/m}^{k,k^{\prime}}(q)$ and $\Psi_{s/t}^{k,k^{\prime}}(q)$ are constant for all $k$ and $k^{\prime}$. 
Since in the second iteration $F_{d/m/s/t}^{k,k^{\prime}}(q)$ and $\Gamma_{d/m/s/t}^{k,k^{\prime}}(q)$ are no longer taking the simple values $(U, -U, 2U, 0)$, the stripes appear in the reducible vertex. 
Though $F_{d/m/s/t}^{k,k^{\prime}}(q)$ and $\Gamma_{d/m/s/t}^{k,k^{\prime}}(q)$ contain structures that strongly deviate from the constant background,
 the only structure of the reducible vertex $\Phi_{d/m}^{k,k^{\prime}}(q)$ and $\Psi_{s/t}^{k,k^{\prime}}(q)$  extending in the Matsubara frequency space  is the stripes.
 Other local structures inside the smaller parameter range (the light-red region), which can be pronounced in some cases, will be treated without kernel approximation. 
Thus, as the first-level approximation, the choice of single-$q$ dependent kernel functions $\tilde{\Phi}_{d/m}(q)$ and $\tilde{\Psi}_{s/t}(q)$ is justified as an approximation for large Matsubara frequencies.
  
Further improvement of this kernel-approximation is possible. For the second level kernel-approximation, we consider kernel functions $\tilde{\Phi}^{k}_{d/m}(q)$ and $\tilde{\Psi}_{s/t}^{k}(q)$ depending on two-arguments, which is in line with the analysis of Ref.~\cite{PhysRevB.86.125114}.
The additional dependence on $k$ in the second-level kernel approximation allows us to also incorporate the crossing stripes of the reducible vertex functions, see Fig.~\ref{Fig:reducible_vertex}(a). 
In practice, we take $\Phi_{d/m}^{k,k^{\prime}}(q)$ and $\Psi_{s/t}^{k,k^{\prime}}(q)$ at one of the edges of the given parameter range, for instance at $k^{\prime}=-30$ in Fig.~\ref{Fig:reducible_vertex}(a), to be the new kernel $\tilde{\Phi}^{k}_{d/m}(q)\approx \Phi_{d/m}^{k,-30}(q)$ and $\tilde{\Psi}_{s/t}^{k}(q)\approx \Psi_{s/t}^{k,-30}(q)$.
The kernel function, in the second-level approximation, is then given as $\tilde{\Phi}_{d/m}^{k}(q)+\tilde{\Phi}_{d/m}^{k^{\prime}}(q) - \tilde{\Phi}_{d/m}(q)$, where $\tilde{\Phi}_{d/m}(q)$ is the first-level kernel function representing the background of the reducible vertex.
Similar expression can be formulated for the particle-particle channel.

The kernel approximations have strong implications for the two problems we discussed before.  
As our numerical study below shows that the open boundary problem of the vertex functions can be efficiently solved by supplementing the reducible vertex functions with the corresponding kernel functions whenever their arguments exceed the parameter space available in the calculations. 
To this end, we illustrate our philosophy of the kernel-approximation in Fig.~\ref{Fig:reducible_vertex}(b), where we show the two different parameter spaces discussed in the beginning of this section as light-blue and light-red squares.     
Only inside the smaller parameter space (light-red square) the full vertex $F_{d/m/s/t}^{k,k^{\prime}}(q)$ can be calculated from the reducible vertex functions $\Phi_{d/m}^{k,k^{\prime}}(q)$ and $\Psi_{s/t}^{k,k^{\prime}}(q)$. 
Outside of the light-red region, in the first-level kernel approximation, the full vertex functions are calculated from $\tilde{\Phi}_{d/m}(q)$ and $\tilde{\Psi}_{s/t}(q)$, or in the second-level kernel approximation from $\tilde{\Phi}_{d/m}^{k}(q)+\tilde{\Phi}_{d/m}^{k^{\prime}}(q) - \tilde{\Phi}_{d/m}(q)$ and   $\tilde{\Psi}_{s/t}^{k}(q)+\tilde{\Psi}_{s/t}^{k^{\prime}}(q) - \tilde{\Psi}_{s/t}(q)$. 
In this way, $F_{d/m/s/t}^{k,k^{\prime}}(q)$ and $\Gamma_{d/m/s/t}^{k,k^{\prime}}(q)$ can be calculated in the full parameter space defined in the calculations. 

\subsection{High-frequency regulation}
To close the self-consistent loop for the parquet equations, the self-energy also needs to be updated. 
As explained before, the sum in Eq.~(\ref{PA_Sigma}) is performed in a finite interval, which corresponds to a truncation of the vertex functions at the boundary. 
Generally, for a sum in a finite interval $(-a,a)$ the truncation effect can only be eliminated when $a$ is large enough so that the quantity to be summed becomes negligibly small at the boundary. 
However, this is not the case for the vertex functions, which extend to infinite values of $k$ and $k^{\prime}$. 
In this section, we show that, based on the two-level kernel approximation introduced above, we can write down auxiliary vertex functions that match the exact complete vertex $F_{d/m/s/t}^{k,k^{\prime}}(q)$ at and beyond the interval boundary. 
Thus their difference becomes zero at the boundary, and can be safely summed over in the finite interval.
As a principle, such an auxiliary function has to be free of the boundary issue, as it is supposed to account for the asymptotics that is not available in the finite parameter space.    

We propose the following  auxiliary function for the full vertex in the density channel, very similar asymptotic functions can be readily formulated for other channels:
\begin{eqnarray}\label{auxiliary_function}
\tilde{F}_{d}^{k,k^{\prime}}(q) =&& U + \tilde{\Phi}_{d}(q) - \frac{1}{2}\tilde{\Phi}_{d}(k^{\prime}-k)-\frac{3}{2}\tilde{\Phi}_{m}(k^{\prime}-k)\nonumber\\
&&+\frac{1}{2}\tilde{\Psi}_{s}(k+k^{\prime}+q)+\frac{3}{2}\tilde{\Psi}_{t}(k+k^{\prime}+q)\:.
\end{eqnarray}
In terms of Fig.~\ref{Fig:reducible_vertex}(b), this is equivalent to calculate $F_{d}^{k,k^{\prime}}(q)$ from the (approximate) kernel functions in both, the smaller and larger, interval.   
Here, for a simple demonstration, Eq.~(\ref{auxiliary_function}) is constructed from the first-level kernel functions.
Similarly, one can also construct this function by using the second-level kernel functions, the resulting auxiliary functions $\tilde{F}_{d}^{k,k^{\prime}}(q)$ will then become a better approximation to the exact complete vertex $F_{d}^{k,k^{\prime}}(q)$.  

Instead of using Eq.~(\ref{PA_Sigma}), with the help of this auxiliary vertex function we now calculate the self-energy as:
\begin{eqnarray}
\Sigma(k) &=&\tilde{\Sigma}(k)-\frac{UT^{2}}{4N}\sum_{k^{\prime},q}G(k+q)G(k^{\prime}+q)G(k^{\prime})\nonumber\\
&&\hspace{1cm}\times[\Delta F_{d}^{k, k^{\prime}}(q)-\Delta F_{m}^{k, k^{\prime}}(q)]\nonumber\\
&&\hspace{0.8cm}-\frac{UT^{2}}{4N}\sum_{k^{\prime},q}G(q-k^{\prime})G(q-k)G(k^{\prime})\nonumber\\
&&\hspace{1cm}\times[\Delta F_{s}^{k, k^{\prime}}(q)+\Delta F_{t}^{k, k^{\prime}}(q)]\:.
\end{eqnarray}
Here, $\Delta F_{d/m/s/t}^{k,k^{\prime}}(q)=F_{d/m/s/t}^{k,k^{\prime}}(q)-\tilde{F}_{d/m/s/t}^{k,k^{\prime}}(q)$; and $\tilde{\Sigma}(k)$ is the self-energy calculated from the kernel functions in all channels. 

In order to faithfully account for full vertex functions at arbitrary $k, k^{\prime}$ and $q$ in $(-\infty, \infty)$, we further split $\tilde{\Sigma}(k)$ into $\tilde{\Sigma}_{1}(k)$ and $\tilde{\Sigma}_{2}(k)$, where $\tilde{\Sigma}_{1}(k)$ contains only the contribution from the $(U, -U, 2U, 0)$ components, while $\tilde{\Sigma}_{2}(k)$ contains the rest of the auxiliary functions [see Eq.~(\ref{auxiliary_function})].   
$\tilde{\Sigma}_{1}(k)$ can then be efficiently calculated as follows
\begin{eqnarray}
\tilde{\Sigma}_{1}(k) &=& -\frac{U^{2}T^{2}}{2N}\sum_{k^{\prime},q}[G(k+q)G(k^{\prime}+q)G(k^{\prime})\nonumber\\
&&\hspace{1.5cm}+G(q-k^{\prime})G(q-k)G(k^{\prime})]\nonumber\\
&=&-U^{2}{\cal FFT}^{-1}[G^{2}(r)G(-r)]\;.
\end{eqnarray}
Here, $G(r)$ is the Fourier component of $G(k)$, and ${\cal FFT}^{-1}$ the (fast) Fourier transformation between these (in this transformation the anti-periodic boundary condition in the imaginary-time space has been taken into account). 
Thus, $\tilde{\Sigma}_{1}(k)$ incorporates the contribution from the lowest-order complete vertex function, {\it i.e.}, the bare Coulomb interaction, for all frequencies and momentum variables. $\tilde{\Sigma}_{1}(k)$ is nothing but the self-energy from the second-order Feynman diagram.
As for $\tilde{\Sigma}_{2}(k)$, we perform the direct sum over $k^{\prime}$ and $q$ in a much larger parameter space which is possible thanks to the kernel approximation. In practice, we usually take this space two or three times larger than the bigger parameter space used for calculating the various vertex functions [the light-blue region in Fig.~\ref{Fig:reducible_vertex}(b)].

The full vertex does not decay asymptotically but extends with finite values to largest $k$, $k^{\prime}$ and $q$. However, due to the  three single-particle propagators $G$ in Eq.~(\ref{PA_Sigma}), the product $GGGF$ still goes to zero asymptotically for large   $k$, $k^{\prime}$ and $q$.
While it is usually difficult for the full vertex functions to work in a large parameter space in practice, this is not a problem for the kernel functions which depend only on one or two  arguments.
Thus, the evaluation of $\tilde{\Sigma}_{2}(k)$ can be carried out in a much larger parameter space. 
We note that the high-frequency regulation explained above is very important for $\tilde{\Sigma}(k)$ to reproduce the asymptotic tail of the self-energy function in frequency space correctly, which is crucial for maintaining the correct high-frequency behavior of the two-particle vertex functions, and vice versa. 

\section{Results}\label{Sec:PA_results}
\subsection{Validation against DMFT} 
In this section, we present numerical results to justify our implementation of the parquet equations and to validate the accuracy of the kernel approximation.  
To this end, we consider the Hubbard model on a 2D square lattice with nearest neighbor hopping $t$ and interaction $U$  at inverse temperature $\beta$.
We solve this model using both the DMFT methodology and the parquet equations at a single-momentum point. 
If not mentioned otherwise the results presented in this section represent the solutions with the second-level kernel function and the high-frequency regulation for the self-energy asymptotics introduced in the previous section. 

\begin{figure}[tbp]
\centering
\includegraphics[width=\linewidth]{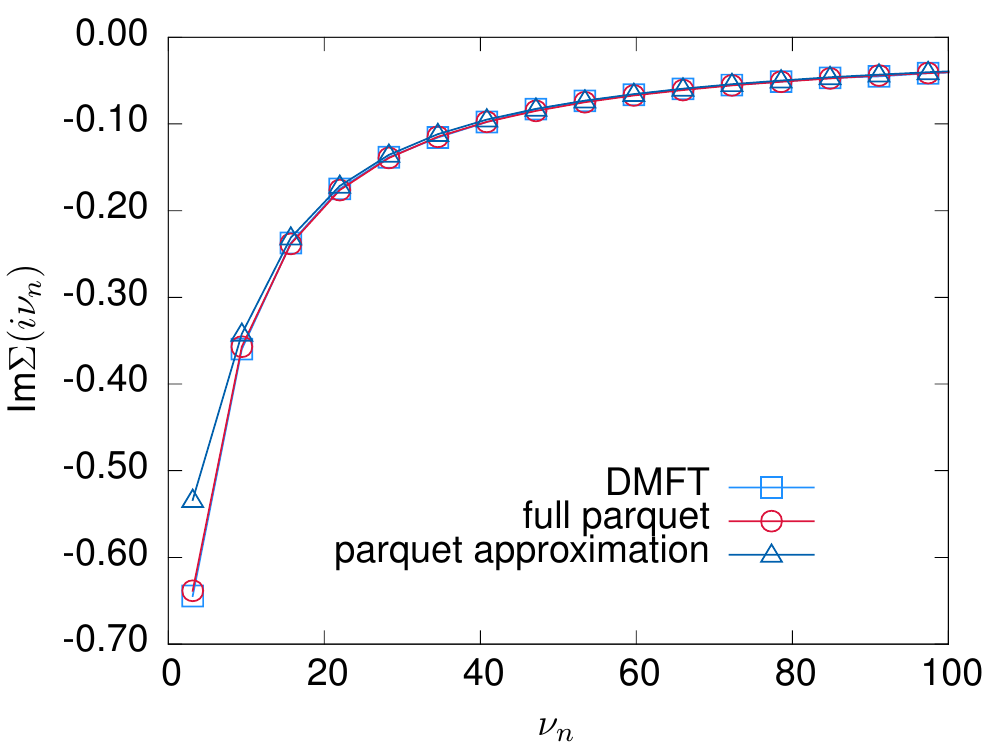}
\caption{Single-particle self-energy  obtained from the parquet equations in the PA and the (local) full parquet calculation employing the kernel approximation. The latter reproduces the DMFT solution with a high precision, but also the PA gives a quite good agreement except for the lowest two Matsubara frequencies.
The parameters for the 2d Hubbard model in DMFT are $\beta = 1$, $U = 4$ (here and in the following $t\equiv 1$). In the parquet equation, 60 Matsubara frequencies have been taken into account in the inner interval of  Fig.~\ref{Fig:reducible_vertex}(b) with the kernel approximation being employed in the outer interval. }
\label{Fig:DMFT_b1u4}
\end{figure}
More specifically, we use  CT-INT as an impurity solver for the DMFT equations, yielding both, the single-particle self-energy and the two-particle vertex function,  in a numerically precise way. 
The DMFT solution provides an unbiased reference for benchmarking our implementation of the parquet equations.  
For a fair comparison, we take the converged DMFT Weiss function ${\cal G}(i\nu_{n})$ as input for the parquet equations. For the other input, {\it i.e.} the fully irreducible vertex function $\Lambda_{d/m/s/t}^{k,k^{\prime}}(q)$, we take two different values:
In one calculation, we take the lowest order  approximation $\Lambda_{d/m/s/t}^{k,k^{\prime}}(q)\approx(U, -U, 2U, 0)$, which corresponds to the PA for the DMFT impurity model. In the other (full parquet) 
calculation we take the CT-INT calculated $\Lambda_{d/m/s/t}^{k,k^{\prime}}(q)$ as input. 
 Since (in contrast to D$\Gamma$A) we do not include a $k$-dependence here,
this calculation exactly reproduces the DMFT results for $F$ and $\Sigma$, if the parquet equations were solved on an infinite frequency interval and statistical errors in CT-INT are negligible. For the given finite frequency interval, this is hence a test for the accuracy of the proposed kernel approximation.

\begin{figure}[tbp]
\centering
\includegraphics[width=\linewidth]{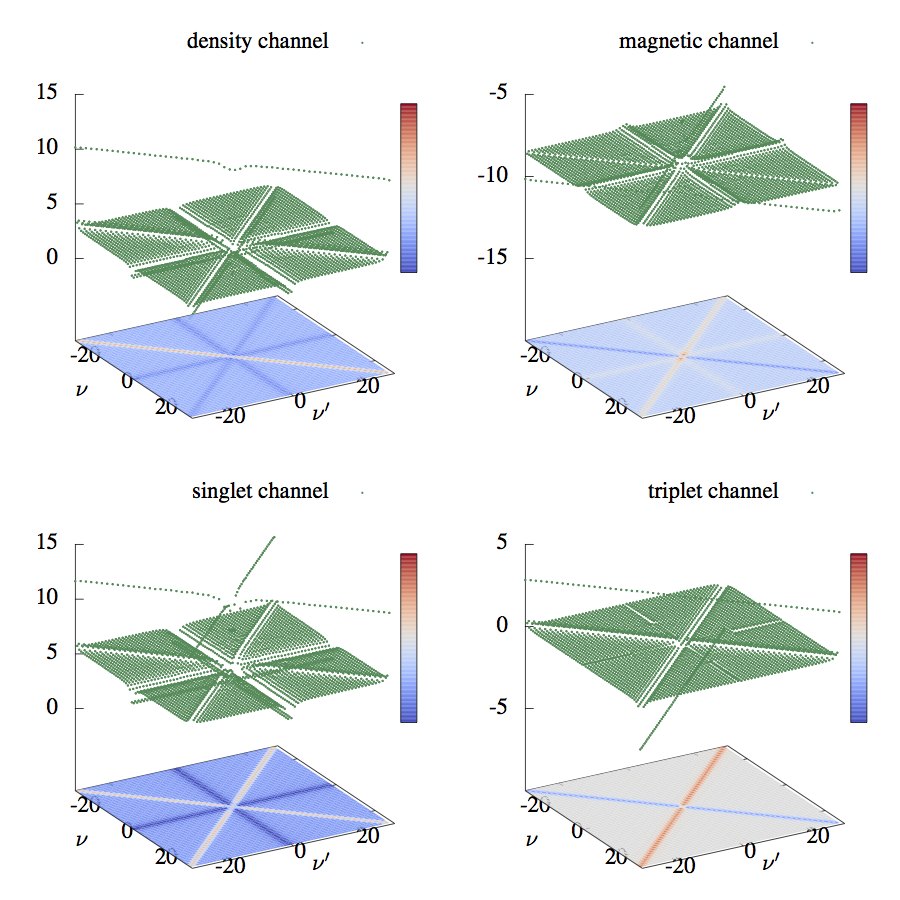}
\caption{Two particle full vertex functions in the four channels as calculated from the parquet equations taking the fully local irreducible vertex from DMFT as an input. The parameters are the same as  in Fig.~\ref{Fig:DMFT_b1u4}.}
\label{Fig:Vertex_DMFTb1u4}
\end{figure}
We show the corresponding full parquet self-energy in Fig.~\ref{Fig:DMFT_b1u4} as empty circles. It nicely reproduces the DMFT solution (empty squares), validating the accuracy of the kernel approximation. Also the PA 
solution (open triangles) agrees well with the DMFT, except for a  small deviation at the first two Matsubara frequencies. 
In particular, the high-frequency tail of the self-energy is nicely reproduced by both parquet solutions. This is an essential check for the algorithm.  
As explained before, a direct truncation of the vertex at the boundary of the available parameter space will lead to the wrong solution of the self-energy, which mainly reflects in the violation of the high-frequency behavior.  

\begin{figure}[tbp]
\centering
\includegraphics[width=\linewidth]{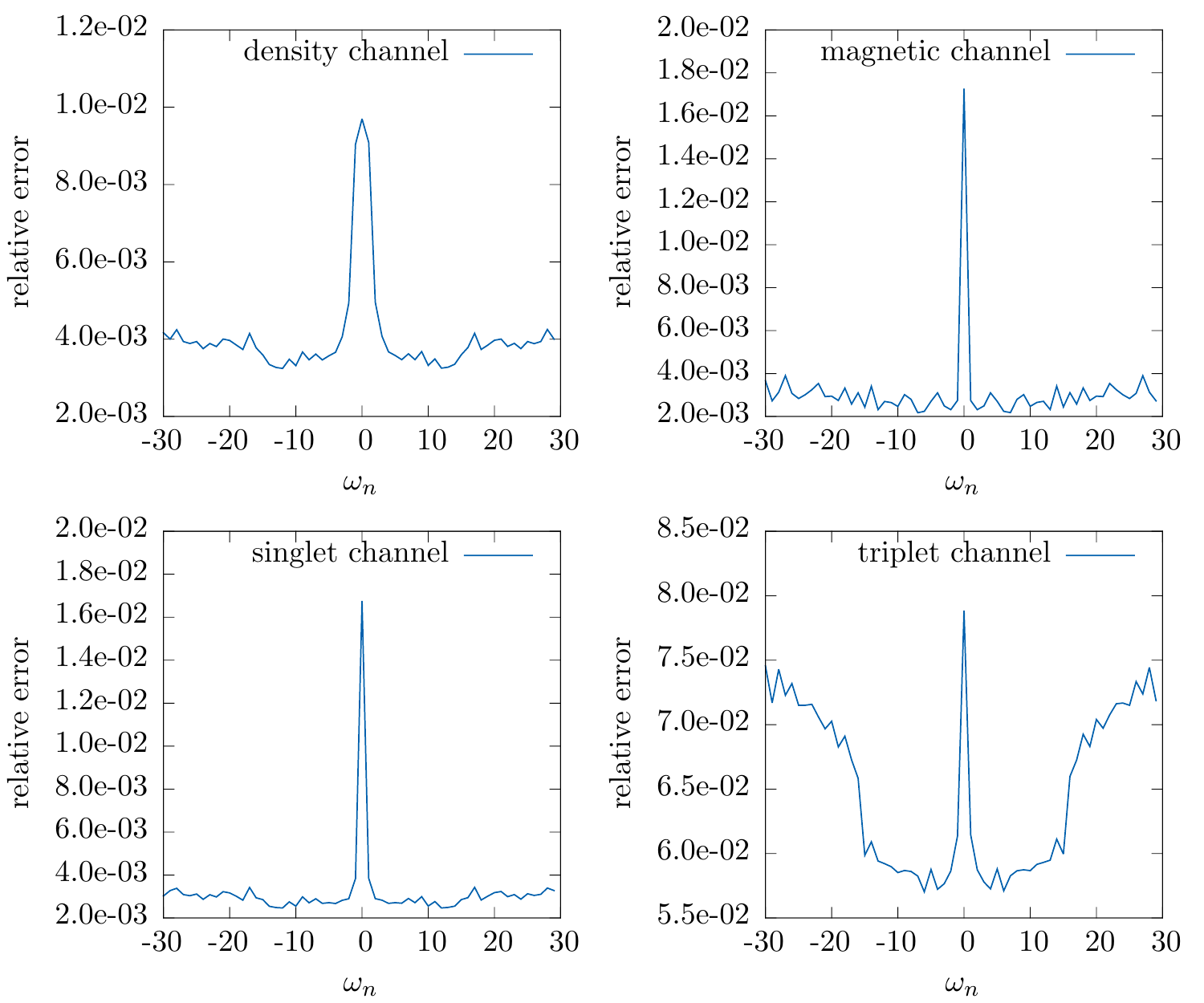}
\caption{The relative error of the complete vertices in Fig.~\ref{Fig:Vertex_DMFTb1u4} with respect to those calculated in DMFT using CT-INT. The relative error is summed up the two fermionic frequencies and is shown as a function of the transfer frequency $\omega_{n}$, see text for more details. Note that this relative error is also subject to the propagation of the statistical error of CT-INT.}
\label{Fig:Vertex_diff}
\end{figure}
Such a violation is a rather common issue appearing in most of the diagrammatic approaches when evaluating the self-energy with only a 
finite numbers of Matsubara frequency. 
In order to achieve a correct high-frequency tail in the self-energy, a few hundreds or even more Matsubara frequencies usually have to be adopted in these approaches~\cite{PhysRevB.55.2122,PhysRevLett.62.961}, which is significantly larger than the number taken in our parquet calculations for similar parameters. 
That is, in all calculations presented in this paper, no more than 60 Matsubara frequencies in each argument are taken, which significantly reduces the demand on the memory for storing all vertex functions. 
Correctly reproducing the high-frequency tail with significantly fewer number of Matsubara frequencies is one of the highlight of our algorithm. 

\begin{figure}[tbp]
\centering
\includegraphics[width=\linewidth]{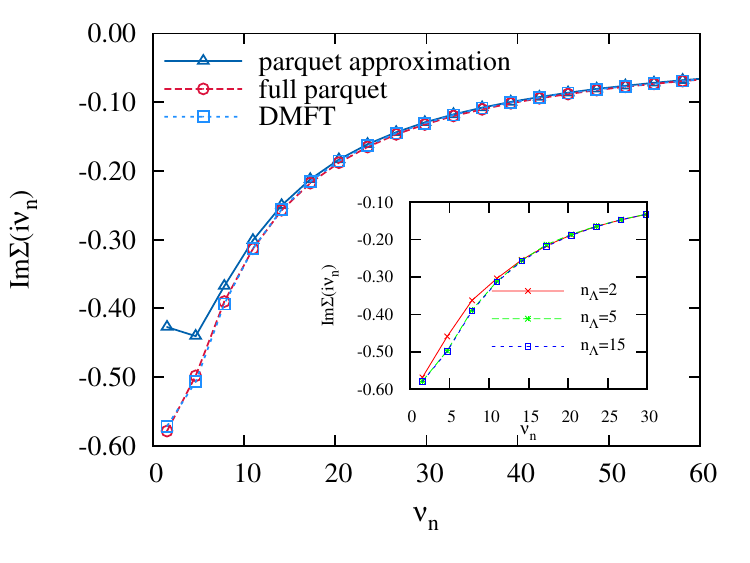}
\caption{Same as Fig.~\ref{Fig:DMFT_b1u4} but for $U=4$ and $\beta = 2$. 
The inset shows the convergence of Im$\Sigma(i\nu_{n})$ with the increase of the frequency cutoff in $\Lambda_{d/m/s/t}$, see the main text for more details.}
\label{Fig:DMFT_b2u4}
\end{figure}

At a lower temperature $\beta = 2$, the full parquet calculation still yields results that excellently agree with the DMFT solution, as shown in Fig.~\ref{Fig:DMFT_b2u4}.
The PA results, on the other hand, deviates more strongly from the DMFT at the low frequencies. This is expected since approximating the fully irreducible vertex by the bare Coulomb interaction is correct only asymptotically for small $U$.
As discussed before, the difference between the PA and the full parquet solutions results from the different values for the fully irreducible vertex function $\Lambda^{k,k^{\prime}}_{d/m/s/t}(q)$ used in the calculations. 
More specificly, in the full parquet calculation, we take $\Lambda_{d/m/s/t}^{k,k^{\prime}}(q)$ obtained from the DMFT (CT-INT) with $31$ Matsubara frequencies for each argument, {\it i.e.} $k, k^{\prime}$ and $q$ are in $[-n_{\Lambda}, n_{\Lambda}]=[-15,15]$, and then extend $\Lambda_{d/m/s/t}^{k,k^{\prime}}(q)$  to $[-30, 30]$ by supplementing it with the lowest order values of these vertices, {\it i.e.} $(U, -U, 2U, 0)$.
In the PA calculations, we take $\Lambda^{k,k^{\prime}}_{d/m/s/t}(q)$ as $(U, -U, 2U, 0)$ everywhere in $[-30, 30]$. 
In order to see the convergence of the full parquet calculation with respect to $n_{\Lambda}$, the inset of Fig.~\ref{Fig:DMFT_b2u4} shows solutions of the full parquet for three different cutoff $n_{\Lambda}$. 
We find a converged solution for $n_{\Lambda}\ge5$. 
As it is known, to obtain the fully irreducible vertex $\Lambda^{k,k^{\prime}}(q)$ with large frequency cutoff is numerically very challenging . 
The inset of Fig.~\ref{Fig:DMFT_b2u4} shows that a relatively small value of cutoff $n_{\Lambda}$ is sufficient to converge the solution (if there exists a convergence) to the correct values. 

Such excellent agreement is not only achieved for the self-energy, we also find that the  full parquet equations give almost identical two-particle vertex functions in all channels (Fig.~\ref{Fig:Vertex_DMFTb1u4}) when compared to the DMFT. 
In Fig.~\ref{Fig:Vertex_diff}, we calculate their relative difference $\sum_{\nu,\nu^{\prime}}|\Delta F^{\nu,\nu^{\prime}}_{d/m/s/t}(\omega)|/|\sum_{\nu,\nu^{\prime}}|F^{{\rm DMFT}, \nu,\nu^{\prime}}_{d/m/s/t}(\omega)|$ by summing up the two fermionic frequencies $\nu, \nu^{\prime}$ and show it as a function of the transfer frequency $\omega_{n}$. 
Here, $\Delta F^{\nu,\nu^{\prime}}_{d/m/s/t}(\omega)=F^{{\rm PARQUET}, \nu,\nu^{\prime}}_{d/m/s/t}(\omega)-F^{{\rm DMFT}, \nu,\nu^{\prime}}_{d/m/s/t}(\omega)$.
The overall amplitude of their differences are small and the biggest deviation appears at $\omega_{n}=0$.
This is expected as, in the reducible vertex,  for any $\nu$ and $\nu^{\prime}$ the largest absolute value is at $\omega_{n}=0$.
It is then easier for an error of the reducible vertex at $\omega_{n}=0$ to propagate to the complete vertex $F^{\nu,\nu^{\prime}}_{d/m/s/t}(\omega_{n})$.
In the triplet channel, we also notice that the relative error is large at larger frequencies, too. 
This is due to the statistical error of the CT-INT and the extrapolation error in the fully localized vertex function $\Lambda_{t}^{\nu,\nu^{\prime}}(\omega)$, which was only calculated up to $|\omega_{n}|=15$ in the CT-INT. 
Let us emphasize that the two-particle vertex $F_{d/m/s/t}^{k,k^{\prime}}(q)$ at larger frequencies are calculated from the kernel-approximation. The small error in this regime, especially in the density, magnetic and singlet channels, shows that the kernel-approximation correctly reproduces the asymptotics of the two-particle vertex functions.    

The agreement in both the single- and two-particle quantities clearly demonstrates that our implementation of the parquet equations fully respects the self-consistency at both singe- and two-particle levels. 
It should be noted that the availability of the two-particle vertex function as output is one of the striking features of the parquet theory. 
The two-particle vertex functions play a crucial role in various diagrammatic approaches~\cite{PhysRevB.75.045118, Held01062008, PhysRevB.77.033101, PhysRevB.79.045133,PhysRevLett.102.206401,doi:10.1143/JPSJ.79.094707,PhysRevB.88.115112,PhysRevB.91.165134}  that construct non-local correlations starting from a local 
DMFT~\cite{RevModPhys.68.13} solution.  
In the dual-fermion (DF)~\cite{PhysRevB.77.033101, PhysRevB.79.045133,PhysRevLett.102.206401},  functional renormalization group enhanced DMFT (DMF$^2$RG)~\cite{PhysRevLett.112.196402}, the non-local expansion (NLE)~\cite{PhysRevB.91.165134} and the three-leg vertex (TRILEX)~\cite{PhysRevB.92.115109} approaches the full vertex functions $F_{d/m/s/t}^{k,k^{\prime}}(q)$ are used to restore the non-local dependence in the self-energy. 
In ladder D$\Gamma$A~\cite{PhysRevB.75.045118, Held01062008} and the one-particle irreducible (1PI) approach~\cite{PhysRevB.88.115112} the channel-dependent irreducible vertex functions $\Gamma_{d/m/s/t}^{k,k^{\prime}}(q)$ are the building blocks for the non-local self-energy diagrams.   Full  parquet D$\Gamma$A~\cite{PhysRevB.91.115115} starts, as we do here, with the most compact and local object, {\it i.e.} the fully irreducible vertex $\Lambda_{d/m/s/t}^{k,k^{\prime}}(q)$.
To obtain these necessary vertex functions is not a trivial task. 
Exact numerical methods, such as quantum Monte Carlo (QMC) or exact diagonalization (ED), are usually employed. We have shown in this paper that, in addition to these approaches, the parquet equations provide another tool that is more flexible than the QMC and ED in many situations, as it can be applied to cases out of half filling, cluster systems, multiorbital materials, {\it etc}. We believe our implementation of the parquet equations smoothes the way for other many-body methods~\cite{PhysRevB.77.033101, PhysRevB.79.045133,PhysRevLett.102.206401, PhysRevLett.112.196402,PhysRevB.91.165134} that are based on the two-particle vertex. 

\begin{figure}
\centering
\includegraphics[width=\linewidth]{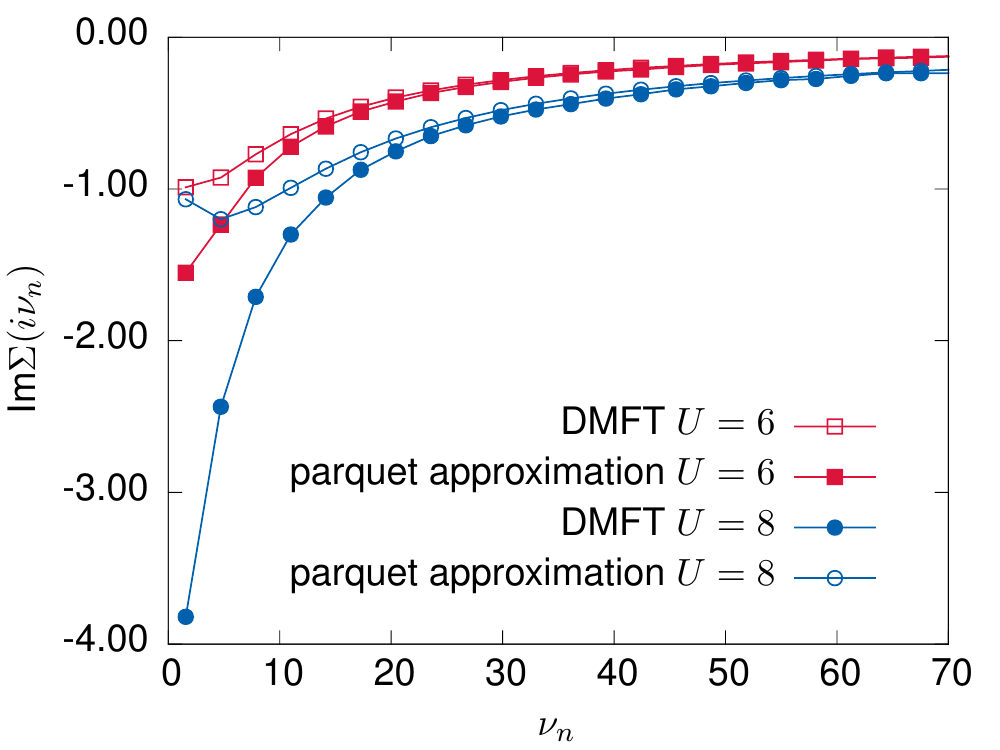}
\caption{With the kernel approximation and high-frequency regulation, the convergence at $U=6$ and $U=8$ can also be achieved in the PA. Here the inverse temperature is the same as in Fig.~\ref{Fig:DMFT_b2u4}. }
\label{Fig:DMFT_b2u6u8}
\end{figure}
Another feature of our parquet implementation is the improved convergence of the algorithm. 
As displayed in Fig.~\ref{Fig:DMFT_b2u6u8}, with the kernel approximation, $U=6$ and $U=8$ can also be converged, which is difficult to achieve in other implementations~\cite{PhysRevE.87.013311, PhysRevE.80.046706}.
The improved convergence is mainly due to the correct understanding of the vertex structure and the subsequently proposed kernel approximation.  
In implementations without auxiliary high-frequency functions, one has to enlarge the frequency range 
to achieve a better convergence. 
However, the rapid growth in the memory demand usually forbids one to do so.    
Comparing Fig.~\ref{Fig:DMFT_b2u6u8} with Fig.~\ref{Fig:DMFT_b2u4} immediately implies that, with the increase of interaction strength, the deviations of the PA from the DMFT become more and more pronounced. 
This, in principle, should be corrected when the full parquet calculations are performed. However,  we noticed that the convergence in the full parquet calculation is generally slower than in the PA, and for these value of interactions, {\it i.e.} $U=6, 8$ and even larger, we did not achieve the convergence in the full parquet calculations, which is mainly due to the almost singular value of $\Lambda_{d/m/s/t}^{k,k^{\prime}}(q)$ occurring at larger values of $U$~\cite{PhysRevLett.110.246405}.

\subsection{Dynamical vertex approximation}
\begin{figure}[tbp]
\centering
\includegraphics[width=\linewidth]{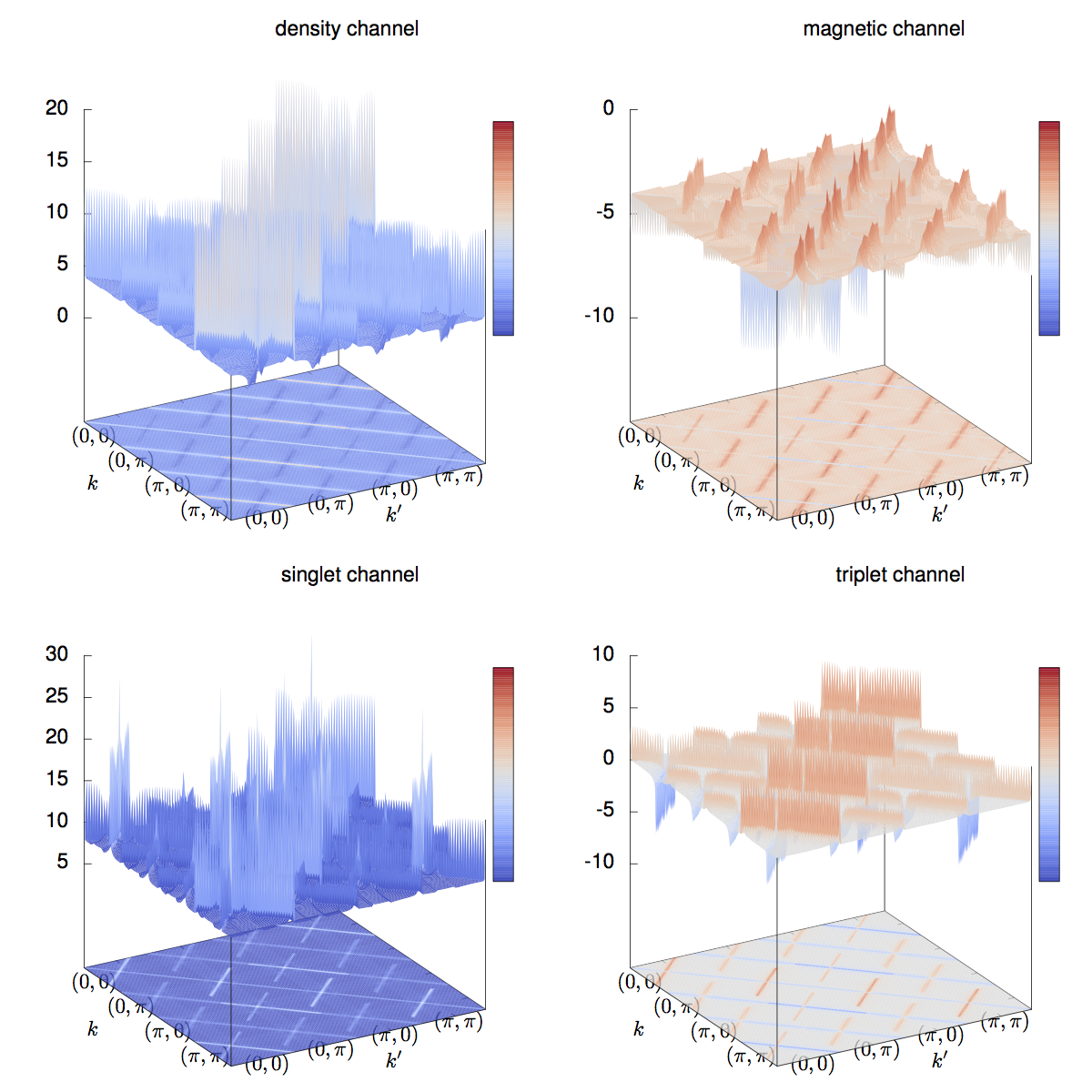}
\caption{Non-local full vertex  obtained by D$\Gamma$A at $\beta t=2, U/t=4$ for a $2\times2$ momentum patch. $F_{d/m/s/t}^{k,k^{\prime}}(q)$ is shown as a function of $k$ and $k^{\prime}$ for fixed $q=0$.}
\label{Fig:Vertex_b2u4}
\end{figure}
In this section, we go beyond the DMFT solution of the Hubbard model discussed in the last section, where the parquet equations are solved without $k$-dependence (for a single $k$ point). Instead we solve the parquet equations for a  $2\times2$ patch-grid in momentum space using the local fully irreducible vertex as an input. 
This is the parquet D$\Gamma$A which includes non-local correlations beyond DMFT~\cite{PhysRevB.91.115115}.
Fig.~\ref{Fig:Vertex_b2u4} shows the non-local, full vertex functions at $\beta =2$ and $U=4$ ($t\equiv 1$) as functions of $k$ and $k^{\prime}$ with $q=0$. 
In each compound index $k$, there are four different momenta, which results in $64$ momentum patches for each vertex function.
Fig.~\ref{Fig:Vertex_b2u4} shows the $16$ patches for $q=0$.   
It is obvious from Fig.~\ref{Fig:Vertex_b2u4} that the full vertex shows a strong momentum dependence, that is also very channel-dependent.
While we here only show results for a $2\times2$ patch-grid, solving
the parquet equations for larger clusters is possible 
owing to the economic use of memory in our kernel approximation.
We found our implementation to be feasible also for calculations on $4\times4$ clusters. Further algorithmic improvements regarding  parallelization and memory management should allow for even larger cluster sizes.

\section{Summary and outlook} \label{Sec:PA_summary}
In this paper, we have proposed a new implementation of the parquet equations and applied it to the one-band Hubbard model in  DMFT and D$\Gamma$A.
We found that it is crucial to respect the correct structure of the vertex functions to simultaneously maintain the self-consistency at both single- and two-particle levels. 
Among the various two-particle vertex functions, the reducible vertex in each channel plays an important role in the parquet equations in the sense that it generates the major structure of the other vertex functions. 
This important observation motivates us to propose a two-level kernel approximation on the reducible vertex $\Phi_{d/m}^{k,k^{\prime}}(q)$ and $\Psi_{s/t}^{k,k^{\prime}}(q)$, which effectively reduces its three-argument dependence to a one/two-argument dependence. 
Employing this two-level kernel approximations in a larger frequency interval, greatly simplifies the calculation.  In particular, it faithfully respects the open boundary condition of the vertex functions in Matsubara frequency space.  Based on the kernel function, we also proposed an auxiliary function to carefully incorporate the high-frequency information missing in the finite sum evaluation of the self-energy.
  
We showed that the two-level kernel approximation and the high-frequency regulation are efficient for solving the parquet equations. 
For the single impurity Anderson model  a very impressive agreement with the DMFT can be achieved which validates our approach. We also demonstrate that the PA works quite well as long as $U$ is not too large. Let us note that 
the kernel approximation and the high-frequency regulation also improve the convergence, which further enhances the applicability of this approach. 
The calculated two-particle vertex functions can be used as a starting point by other many-body approaches, such as the ladder-D$\Gamma$A, 1PI approach, DMF$^{2}$RG, DF, NLE and TRILEX.

The proposed two-level kernel approximations and the high-frequency regulations are compatible with the PA and the full parquet D$\Gamma$A which we were able to perform, for the first time,  in two dimensions. Physically, the advantage over previously employed ladder D$\Gamma$A~\cite{ PhysRevB.91.125109, PhysRevLett.107.256402} is that in the full parquet  D$\Gamma$A also the particle-particle (Cooper) channel is included. This allows to study spin-fluctuation mediated superconductivity~\cite{PhysRevLett.62.961, Scalapino_1} and instabilities towards stripe phases~\cite{PhysRevB.40.7391}. 
Let us note that non-local interactions can also be included straightforwardly. For example, studying an extended Hubbard model with nearest-neighbor interaction and the competition between the long-range magnetic and charge instabilities is possible.

\acknowledgments
We want to thank S. Andergassen, G. Rohringer and A. Toschi for the valuable comments. We acknowledge support from  European Research Council under the European Union's Seventh Framework Programme (FP/2007-2013)/ERC through grant agreement n. 306447 ({\em AbinitioD$\rm \Gamma$A}). We grateful for the hospitality of the Aspen Center for Physics, which is supported by National Science Foundation grant PHY-1066293. The computational results presented have been achieved using the Vienna Scientific Cluster (VSC).


\appendix

\section{formulation of the parquet equations}\label{Sec:PA_formalism}
In this section, we present the necessary notations that are used  in this paper.  Based on these notations, the parquet equations are derived under $SU(2)$ symmetry. 
The complete derivation of the parquet formulation concerns two parts: the coupled equations for the two-particle vertex functions in all channels and the one-particle self-energy. 

Throughout this paper, we will consider the half-filled single-band Hubbard model on a square lattice and used its DMFT solution as a benchmark for testing the numerical feasibility of our approach. 
The Hubbard Hamiltonian reads
\begin{equation}\label{Hubbard}
H=\sum_{\mathbf{k},\sigma}\epsilon_{\mathbf{k}}c_{\mathbf{k},\sigma}^{\dagger}c_{\mathbf{k},\sigma}+ U\sum_{i}n_{i\uparrow}n_{i\downarrow}\;.
\end{equation}
Here, $\mathbf{k}$ represents a momentum vector in the two-dimensional (2D) square lattice,  $\epsilon_{\mathbf{k}}=-2t (\cos k_x+\cos k_y)$, $c_{\mathbf{k},\sigma}^{\dagger}$ ($c_{\mathbf{k},\sigma}$) creates (annihilates) an electron with momentum $\mathbf{k}$ and spin $\sigma\in\{\uparrow,\downarrow\}$, and  $n_{i\sigma}\equiv c_{i,\sigma}^{\dagger}c_{i,\sigma}$ is the number operator on lattice site $i$.  

\subsection{Notations}
First, we introduce the definition for the two-particle susceptibility $\chi$, from which other vertex functions can be derived. 
The particle-hole and particle-particle susceptibilities are defined as
\begin{subequations}
\begin{align}
\chi_{ph,\sigma\sigma^{\prime}}^{k,k^{\prime}}(q) 
=& \sum_{ijkl} e^{-ikr_{i}}e^{i(k+q)r_{j}}e^{-i(k^{\prime}+q)r_{k}}e^{ik^{\prime}r_{l}} \nonumber\\
&\times \langle T_{\tau}c_{\sigma}^{\dagger}(r_{i})c_{\sigma}(r_{j}) c^{\dagger}_{\sigma^{\prime}}(r_{k})c_{\sigma^{\prime}}(r_{l})\rangle\;, \\
\chi_{pp,\sigma\sigma^{\prime}}^{k,k^{\prime}}(q) 
=&\sum_{ijkl}e^{-ikr_{i}}e^{i(q-k^{\prime})r_{j}}e^{-i(q-k)r_{k}}e^{ik^{\prime}r_{l}}\nonumber\\
&\times\langle T_{\tau}c_{\sigma}^{\dagger}(r_{i})c_{\sigma}(r_{j})c_{\sigma^{\prime}}^{\dagger}(r_{k})c_{\sigma^{\prime}}(r_{l})\rangle\;.
\end{align}
\end{subequations}
Here, $r=(\mathbf{r},\tau)$ with lattice site $\mathbf{r}$ and imaginary time $\tau$,  $k=(\mathbf{k},i \nu)$ with wave vector $\mathbf{k}$ and Matsubara frequency $i \nu$, 
 and $q=(\mathbf{q}, i\omega)$ with the transfer momentum and bosonic frequency. 
$\sum_{ijkl}$ shall be understood as $T\sum_{\mathbf{r}_{i}\cdots\mathbf{r}_{l}}\int_{0}^{\beta}d\tau_{i}\cdots d\tau_{l}$ where $T$ is the temperature. 
Note that the particle-hole and particle-particle excitations are encoded in the same four-point correlator in the above equation, thus, $\chi_{ph,\sigma\sigma^{\prime}}^{k,k^{\prime}}(q)$ and $\chi_{pp,\sigma\sigma^{\prime}}^{k,k^{\prime}}(q)$ are not independent but relate to each other by means of a frequency shift. That is, they are related to each other as $\chi_{pp,\sigma\sigma^{\prime}}^{k,k^{\prime}}(q)=\chi_{ph,\sigma\sigma^{\prime}}^{k,k^{\prime}}(q-k-k^{\prime})$.
The same relation also holds for the complete vertex $F$ and fully irreducible vertex $\Lambda$.

\begin{figure}[tbp]
\centering
\includegraphics[width=0.8\linewidth]{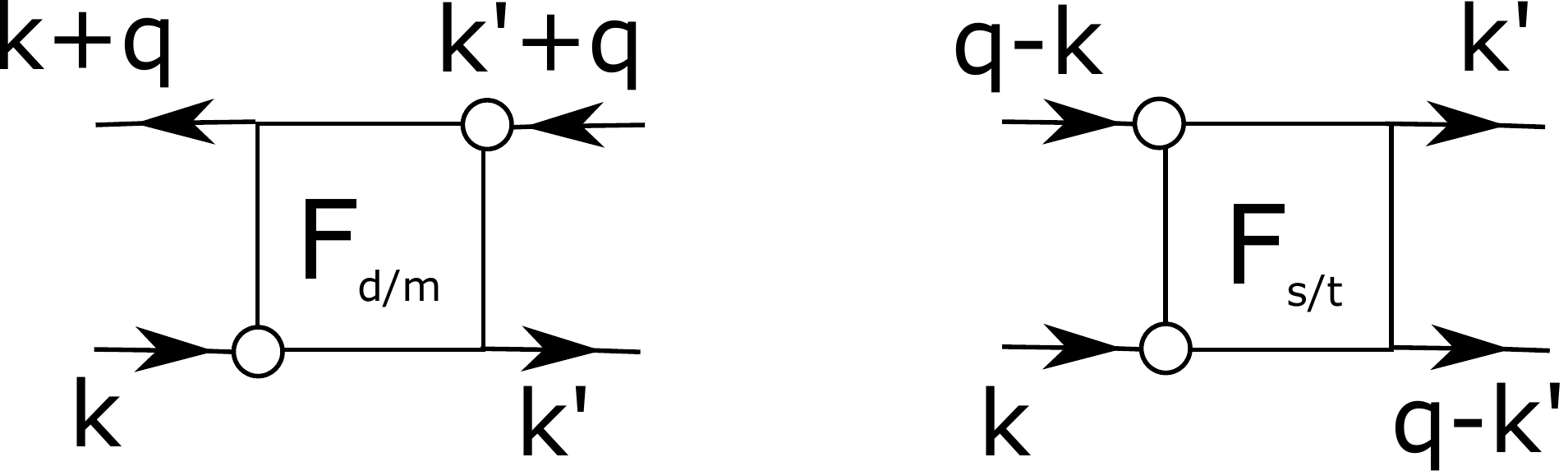}
\caption{Graphical representation of the vertex functions in the particle-hole ($d/m$) and the particle-particle ($s/t$) channels, which apply to all the vertex in this work.}
\label{Fig:2P_notation}
\end{figure}

From the susceptibilities $\chi_{ph,\sigma\sigma^{\prime}}^{k,k^{\prime}}(q)$ and $\chi_{pp,\sigma\sigma^{\prime}}^{k,k^{\prime}}(q)$, the complete (full) vertex functions $F_{ph,\sigma\sigma^{\prime}}^{k,k^{\prime}}(q)$ and $F_{pp,\sigma\sigma^{\prime}}^{k,k^{\prime}}(q)$ can be easily obtained as
\begin{subequations}
\begin{align}
F_{ph,\sigma\sigma^{\prime}}^{k,k^{\prime}}(q)& = -\frac{\chi_{ph,\sigma\sigma^{\prime}}^{k,k^{\prime}}(q)-\chi^{0,kk^{\prime}}_{ph,\sigma\sigma^{\prime}}(q)}{G^{k}_{\sigma}G^{k+q}_{\sigma}G^{k^{\prime}}_{\sigma^{\prime}}G^{k^{\prime}+q}_{\sigma^{\prime}}}\;,\\
F_{pp,\sigma\sigma^{\prime}}^{k,k^{\prime}}(q) &= -\frac{\chi_{pp,\sigma\sigma^{\prime}}^{k,k^{\prime}}(q)-\chi^{0,kk^{\prime}}_{pp,\sigma\sigma^{\prime}}(q)}{G^{k}_{\sigma}G^{q-k^{\prime}}_{\sigma}G^{k^{\prime}}_{\sigma^{\prime}}G^{q-k}_{\sigma^{\prime}}}\;,
\end{align}
\end{subequations}
with the bare bubble susceptibilities $\chi^{0,kk^{\prime}}_{ph,\sigma\sigma^{\prime}}(q) = \frac{\beta}{N}[G_{\sigma}^{k}G_{\sigma^{\prime}}^{k^{\prime}}\delta_{q,0}-G_{\sigma}^{k}G_{\sigma}^{k+q}\delta_{k,k^{\prime}}\delta_{\sigma\sigma^{\prime}}]$ and
$\chi^{0,kk^{\prime}}_{pp,\sigma\sigma^{\prime}}(q) = \frac{\beta}{N}[G_{\sigma}^{k}G_{\sigma^{\prime}}^{k^{\prime}}\delta_{k,q-k^{\prime}}-G_{\sigma}^{k}G_{\sigma}^{q-k}\delta_{k,k^{\prime}}\delta_{\sigma\sigma^{\prime}}]$.
Under the $SU(2)$ symmetry, the full vertex functions (including also the other vertex functions) with different spin configurations can be cast into a more compact form in the density ($d$), magnetic ($m$), singlet ($s$) and triplet ($t$) channels, see Fig.\ \ref{Fig:2P_notation}:
\begin{subequations}
\begin{align}
F_{d/m}^{k,k^{\prime}}(q) = F_{ph,\uparrow\uparrow}^{k,k^{\prime}}(q) \pm F_{ph,\uparrow\downarrow}^{k,k^{\prime}}(q)\;,\\
F_{t/s}^{k,k^{\prime}}(q) = F_{pp,\uparrow\downarrow}^{k,k^{\prime}}(q) \pm F_{pp,\overline{\uparrow\downarrow}}^{k,k^{\prime}}(q)\;.
\end{align}
\end{subequations}

In each channel, the full vertex function can be further decomposed into the two-particle irreducible vertex ($\Gamma_{d/m/s/t}$) and reducible vertex ($\Phi_{d/m}$, $\Psi_{t/s}$) through the Bethe-Salpeter equation (BSE), which
has been thoroughly discussed in many works, see {\it e.g.}~\cite{doi:10.1142/S021797929100016X, Georgphdthesis}. Here, we will  only recall the BSE formulas as used in the derivation of the parquet equations:
\begin{subequations}\label{PA_F_G}
\begin{align}
F_{d/m}^{k,k^{\prime}}(q) = \Gamma_{d/m}^{k,k^{\prime}}(q) + \Phi_{d/m}^{k,k^{\prime}}(q)\;,\\
F_{t/s}^{k,k^{\prime}}(q) = \Gamma_{t/s}^{k,k^{\prime}}(q) + \Psi_{t/s}^{k,k^{\prime}}(q)\;,
\end{align}
\end{subequations}
where the reducible vertex functions depend on the irreducible and full vertex as follows,
\begin{subequations}\label{PA_F_Phi}
\begin{align}
\Phi_{d/m}^{k,k^{\prime}}(q) &= \frac{T}{N}\sum_{k^{\prime\prime}}\Gamma^{k,k^{\prime\prime}}_{d/m}(q)G(k^{\prime\prime})G(k^{\prime\prime}+q)F^{k^{\prime\prime},k^{\prime}}_{d/m}(q)\;,\\
\Psi_{t/s}^{k,k^{\prime}}(q) &= \pm\frac{T}{2N}\sum_{k^{\prime\prime}}\Gamma^{k,k^{\prime\prime}}_{t/s}(q)G(k^{\prime\prime})G(q-k^{\prime\prime})F^{k^{\prime\prime},k^{\prime}}_{t/s}(q)\;.
\end{align}
\end{subequations} 

\subsection{Derivation of the parquet equations}

\begin{figure}[tbp]
\centering
\includegraphics[width=\linewidth]{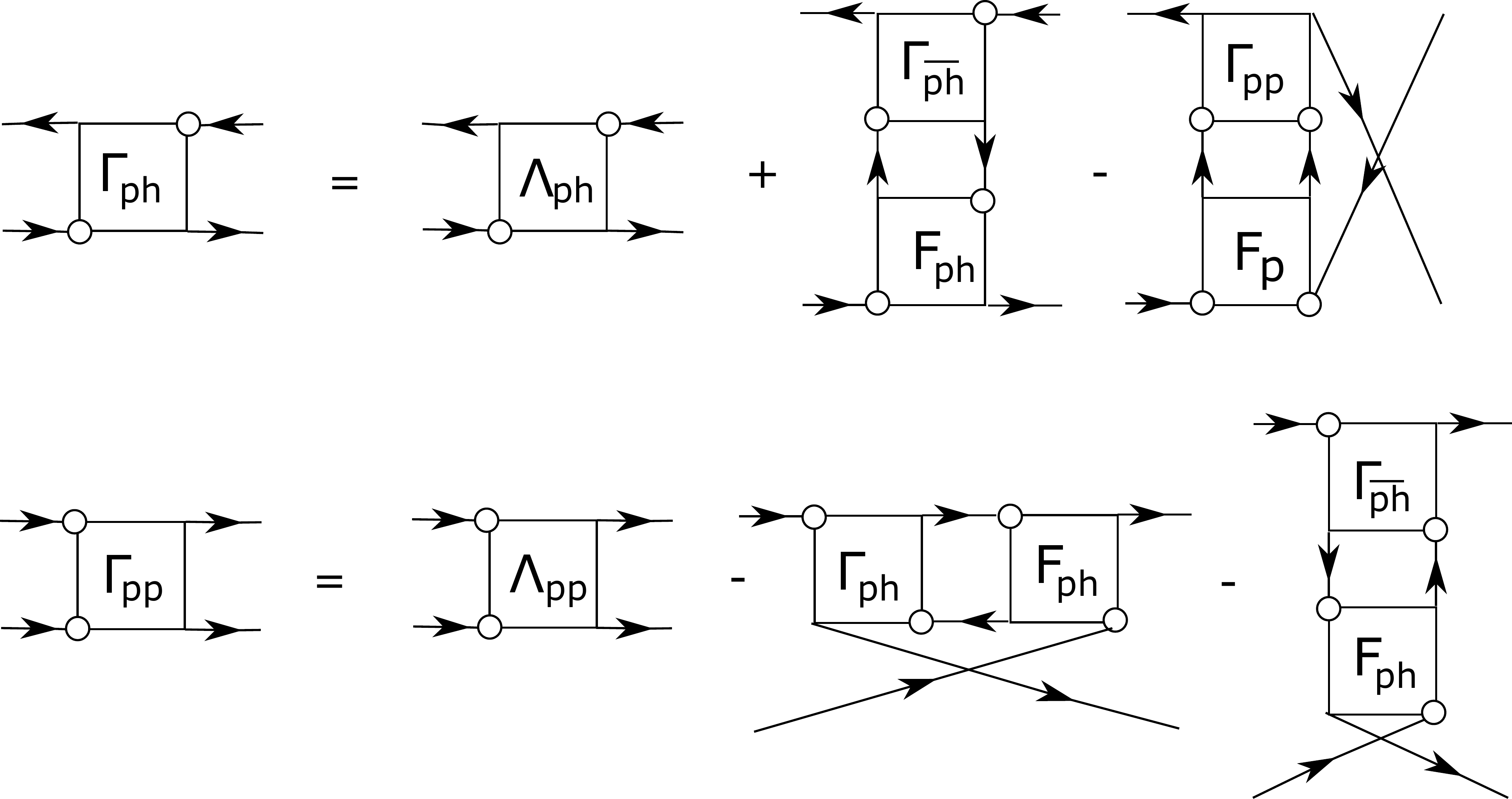}
\caption{Coupled diagrams for the parquet equations in the particle-hole and particle-particle channels. Here the corresponding diagrams in the particle-hole transverse channel has been omitted as it does not lead to independent contributions to the parquet equations and can be derived from the particle-hole channel.}
\label{Fig:parquet_1}
\end{figure}
With the above notations and definitions, we now proceed to derive the parquet equations. 
The irreducible vertex $\Gamma_{d/m/s/t}$ is only irreducible in given channel, while it becomes reducible in other channels.
$\Lambda_{d/m/s/t}$, as the most fundamental one among all vertex functions, is fully irreducible in all channels.
Given $\Lambda_{d/m/s/t}$, the full vertex $F_{d/m/s/t}$, the channel-dependent irreducible vertex $\Gamma_{d/m/s/t}$ and the reducible vertices $\Phi_{d/m}, \Psi_{s/t}$ can be readily calculated from the parquet equation, as represented graphically in Fig.~\ref{Fig:parquet_1}. 
The parquet equation is nothing but a classification of diagrams in terms of their two-particle irreducibility.
Mathematically, by taking the spin-dependence of each diagram into account, we obtain the parquet equation in the particle-hole channel as
\begin{subequations}
\begin{align}
\Gamma_{ph,\uparrow\uparrow}^{k,k^{\prime}}(q)&=\Lambda_{ph,\uparrow\uparrow}^{k,k^{\prime}}(q) + \Phi_{\overline{ph},\uparrow\uparrow}^{k,k^{\prime}}(q)\nonumber\\
&\hspace{1.8cm}-\Psi_{pp,\uparrow\uparrow}^{k,k+q}(k+k^{\prime}+q)\label{PA_Ph1}\;,\\
\Gamma_{ph,\uparrow\downarrow}^{k,k^{\prime}}(q)&=\Lambda_{ph,\uparrow\downarrow}^{k,k^{\prime}}(q) + \Phi_{\overline{ph},\uparrow\downarrow}^{k,k^{\prime}}(q)\nonumber\\
&\hspace{1.8cm}-\Psi_{pp,\overline{\uparrow\downarrow}}^{k,k+q}(k+k^{\prime}+q)\;.\label{PA_Ph2}
\end{align}
\end{subequations} 
After applying the following crossing relations~\cite{Georgphdthesis} 
\begin{subequations}
\begin{align}
\Phi_{\overline{ph},\uparrow\uparrow}^{k,k^{\prime}}(q)&= -\Phi_{ph,\uparrow\uparrow}^{k,k+q}(k^{\prime}-k)\;, \\
\Phi_{\overline{ph},\uparrow\downarrow}^{k,k^{\prime}}(q) &= -\Phi_{m}^{k,k+q}(k^{\prime}-k)\label{CR_1}\;,\\
\Psi_{pp,\uparrow\uparrow}^{k,k^{\prime}}(q) &= \Psi_{t}^{k,k^{\prime}}(q)
=-\Psi_{t}^{k,q-k^{\prime}}(q)\;, \\
\Psi_{pp,\overline{\uparrow\downarrow}}^{k,k^{\prime}}(q) &= -\Psi_{pp,\uparrow\downarrow}^{k,q-k^{\prime}}(q)\;.\label{CR_2}
\end{align}
\end{subequations}
to Eqs.~(\ref{PA_Ph1}) and (\ref{PA_Ph2}), we have 
\begin{subequations}
\begin{align}
\Gamma_{ph,\uparrow\uparrow}^{k,k^{\prime}}(q)&=\Lambda_{ph,\uparrow\uparrow}^{k,k^{\prime}}(q) -\Phi_{ph,\uparrow\uparrow}^{k,k+q}(k^{\prime}-k)\nonumber\\
&\hspace{1.9cm}+\Psi_{t}^{k,k^{\prime}}(k+k^{\prime}+q)\:;\\
\Gamma_{ph,\uparrow\downarrow}^{k,k^{\prime}}(q)&=\Lambda_{ph,\uparrow\downarrow}^{k,k^{\prime}}(q) -\Phi_{m}^{k,k+q}(k^{\prime}-k)\nonumber\\
&\hspace{1.9cm}+\Psi_{pp,\uparrow\downarrow}^{k,k^{\prime}}(k+k^{\prime}+q)\:,
\end{align}
\end{subequations}
which can be equivalently written in the density and magnetic channels as 
\begin{subequations}\label{PA_Gamma_dm}
\begin{align}
\Gamma_{d}^{k,k^{\prime}}(q)=\Lambda_{d}^{k,k^{\prime}}(q) &- \frac{1}{2}\Phi_{d}^{k,k+q}(k^{\prime}-k)\nonumber\\
&- \frac{3}{2}\Phi_{m}^{k,k+q}(k^{\prime}-k)\nonumber\\
&+\frac{1}{2}\Psi_{s}^{k,k^{\prime}}(k+k^{\prime}+q)\nonumber\\
&+\frac{3}{2}\Psi_{t}^{k,k^{\prime}}(k+k^{\prime}+q)\;;\\
\Gamma_{m}^{k,k^{\prime}}(q)=\Lambda_{m}^{k,k^{\prime}}(q) &- \frac{1}{2}\Phi_{d}^{k,k+q}(k^{\prime}-k) \nonumber\\
&+ \frac{1}{2}\Phi_{m}^{k,k+q}(k^{\prime}-k)\nonumber\\
& -\frac{1}{2}\Psi_{s}^{k,k^{\prime}}(k+k^{\prime}+q)\nonumber\\
&+\frac{1}{2}\Psi_{t}^{k,k^{\prime}}(k+k^{\prime}+q)\:. 
\end{align}
\end{subequations}
Similarly, for the particle-particle channel in Fig.~\ref{Fig:parquet_1}, the equations read
\begin{subequations}
\begin{align}
\Gamma_{pp,\uparrow\downarrow}^{k,k^{\prime}}(q)&=\Lambda_{pp,\uparrow\downarrow}^{k,k^{\prime}}(q)-\Phi_{ph,\overline{\uparrow\downarrow}}^{k,q-k^{\prime}}(k^{\prime}-k)\nonumber\\
&\hspace{1.8cm}-\Phi_{\overline{ph},\overline{\uparrow\downarrow}}^{k,q-k^{\prime}}(k^{\prime}-k)\:;\\
\Gamma_{pp,\overline{\uparrow\downarrow}}^{k,k^{\prime}}(q)&=\Lambda_{pp,\overline{\uparrow\downarrow}}^{k,k^{\prime}}(q)-\Phi_{ph,\uparrow\downarrow}^{k,q-k^{\prime}}(k^{\prime}-k)\nonumber\\
&\hspace{1.8cm}-\Phi_{\overline{ph},\uparrow\downarrow}^{k,q-k^{\prime}}(k^{\prime}-k)\:.
\end{align}
\end{subequations}
To simply these equations, we need again Eq.~(\ref{CR_1}) and the following relation:
\begin{subequations}
\begin{align}
\Phi_{\overline{ph},\overline{\uparrow\downarrow}}^{k,k^{\prime}}(q)&=-\Phi_{ph,\uparrow\downarrow}^{k,k+q}(k^{\prime}-k)\:;\\
\Phi_{ph,\overline{\uparrow\downarrow}}^{k,k^{\prime}}(q)&=\Phi_{m}^{k,k^{\prime}}(q)\:.
\end{align}
\end{subequations}
The parquet equations for the particle-particle channel are then found to be:
\begin{subequations}\label{PA_Gamma_st}
\begin{align}
\Gamma_{s}^{k,k^{\prime}}(q)=\Lambda_{s}^{k,k^{\prime}}(q) & 
 +\frac{1}{2}\Phi_{d}^{k,q-k^{\prime}}(k^{\prime}-k)\nonumber\\
 &-\frac{3}{2}\Phi_{m}^{k,q-k^{\prime}}(k^{\prime}-k)\nonumber\\
& +\frac{1}{2}\Phi_{d}^{k,k^{\prime}}(q-k-k^{\prime})\nonumber\\
& - \frac{3}{2}\Phi_{m}^{k,k^{\prime}}(q-k-k^{\prime})\:;\\
\Gamma_{t}^{k,k^{\prime}}(q) = \Lambda_{t}^{k,k^{\prime}}(q)& -\frac{1}{2}\Phi_{d}^{k,q-k^{\prime}}(k^{\prime}-k)\nonumber\\
& -\frac{1}{2}\Phi_{m}^{k,q-k^{\prime}}(k^{\prime}-k)\nonumber\\
&+\frac{1}{2}\Phi_{d}^{k,k^{\prime}}(q-k-k^{\prime})\nonumber\\
&+ \frac{1}{2}\Phi_{m}^{k,k^{\prime}}(q-k-k^{\prime})\:.  
\end{align}
\end{subequations}

\subsection{Crossing symmetry}
An important symmetry that the parquet equations satisfy but that is violated in the Baym-Kadanoff formalism is the crossing symmetry, which for the full vertex reads:
\begin{subequations} 
\begin{align}
F_{d}^{k,k^{\prime}}(q)=&\hspace{0.5cm}\frac{1}{2}F_{s}^{k,k^{\prime}}(k+k^{\prime}+q) + \frac{3}{2}F_{t}^{k,k^{\prime}}(k+k^{\prime}+q)\:;\nonumber\\
F_{m}^{k,k^{\prime}}(q)=&-\frac{1}{2}F_{s}^{k,k^{\prime}}(k+k^{\prime}+q) + \frac{1}{2}F_{t}^{k,k^{\prime}}(k+k^{\prime}+q)\:;\nonumber\\
F_{s}^{k,k^{\prime}}(q)=&\hspace{0.5cm}\frac{1}{2}F_{d}^{k,k^{\prime}}(q-k-k^{\prime})-\frac{3}{2}F_{m}^{k,k^{\prime}}(q-k-k^{\prime})\:;\nonumber\\
F_{t}^{k,k^{\prime}}(q)=&\hspace{0.5cm}\frac{1}{2}F_{d}^{k,k^{\prime}}(q-k-k^{\prime})+\frac{1}{2}F_{m}^{k,k^{\prime}}(q-k-k^{\prime})\:.
\end{align}
\end{subequations}
These equations can be easily verified in the parquet equation~(\ref{PA_F}) by substituting Eq.~(\ref{PA_Gamma_dm}) and (\ref{PA_Gamma_st}) into Eq.~(\ref{PA_F_G}).
A correct solution of the parquet equations certainly should respect this symmetry. 
It has been understood that the above crossing symmetry can be explicitly enforced at each self-consistent step by solving the parquet equations for the full vertex $F_{d/m/s/t}$, {\it i.e.} Eq.\ (\ref{PA_F}), instead of those for $\Gamma_{d/m/s/t}$~\cite{PhysRevE.87.013311}. 
We note that a similar crossing symmetry also applies to the fully irreducible vertex $\Lambda_{d/m/s/t}$.

\subsection{Self-energy from the full vertex}
\begin{figure}[htbp]
\vspace{0.5cm}
\centering
\includegraphics[width=\linewidth]{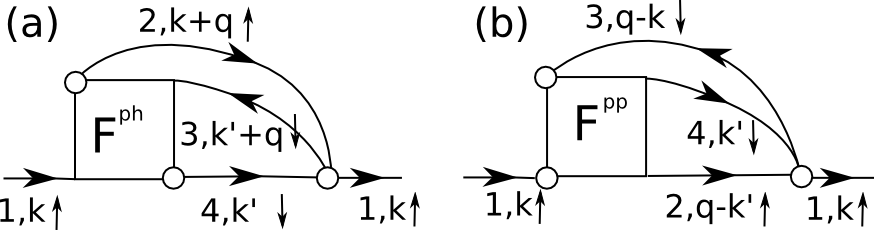}
\caption{The Feynman diagram for the self-energy, which contains  the contributions from both, particle-hole and particle-particle, channels.}
\label{Fig:SD}
\end{figure}
To close the self-consistent loop in the parquet theory, we also need to connect the two-particle full vertex functions $F_{d/m/s/t}$ with the single-particle self-energy $\Sigma(k)$, which is graphically shown in Fig.~\ref{Fig:SD}. This connection can be derived through the Heisenberg equation of motion and is also known as the Schwinger-Dyson equation.
In this context it reads
\begin{eqnarray}\label{PA_Sigma}
\Sigma(k) &=&-\frac{UT^{2}}{4N}\sum_{k^{\prime},q}G(k+q)G(k^{\prime}+q)G(k^{\prime})\nonumber\\
&&\hspace{2cm}\times[F_{d}^{k, k^{\prime}}(q)-F_{m}^{k, k^{\prime}}(q)]\nonumber\\
&&-\frac{UT^{2}}{4N}\sum_{k^{\prime},q}G(q-k^{\prime})G(q-k)G(k^{\prime})\nonumber\\
&&\hspace{2cm}\times[F_{s}^{k, k^{\prime}}(q)+F_{t}^{k, k^{\prime}}(q)]\:.
\end{eqnarray}
Here, the sum over $k^{\prime}$ and $q$ should be done over all Matsubara frequencies. 
In principle, the Hartree and Fock terms need to be added to Eq.~(\ref{PA_Sigma}), but are not relevant for the one-band Hubbard model in the paramagnetic phase.

From $\Sigma$ in turn, the Green function is obtained through the  Dyson equation, which for the sake of completeness reads
 \begin{equation}
   G(k)=[i \omega - \epsilon_{\mathbf k} - \Sigma(k)]^{-1}.
 \label{PA_Dyson}
\end{equation}
This Green function enters Eq.\ (\ref{PA_F_Phi}) which closes the set of equations in the parquet formalism.
\bibliography{ref}

\end{document}